\documentclass[fleqn,10pt]{olplainarticle}

\usepackage{hyperref}
\usepackage{placeins}
\usepackage{float}
\hypersetup{
    colorlinks=true,
    linkcolor=blue,
    citecolor=blue,
    urlcolor=blue,
}
\usepackage[
    style=ieee, 
    citestyle=numeric-comp, 
    sorting=none,  
    backend=biber, 
    url=false, 
    doi=true, 
    isbn=false
]{biblatex}
\title{How many mutations does it take to evolve a new function?} 

\author[1,*]{Nida Farheen}
\author[2]{Zachary Boroda}
\author[3]{Ethan Levien}
\author[4,*]{Jan\'e Kondev}
\affil[1]{Department of Biochemistry, Brandeis University, Waltham, MA; current affiliation: School of Civil and Environmental Engineering, Cornell University, Ithaca, NY}
\affil[2]{Department of Computer Science, Brandeis University, Waltham, MA; current affiliation: Amazon Web Services, New York, NY}
\affil[3]{Department of Mathematics, Dartmouth College, Hanover, NH}
\affil[4]{Department of Physics, Brandeis University, Waltham, MA}
\affil[*]{Correspondence: fn87@cornell.edu; kondev@brandeis.edu}

\keywords{genotype-phenotype maps, Boolean networks, timescale of evolution}

\begin{abstract}
Biochemical and genetic networks in cells evolve to produce new functions. Evolution is guided by the genotype to phenotype map which describes how mutations at the molecular scale lead to changes in network function at the cellular scale. The number of genotypes is astronomically large which makes these maps very high dimensional. In this context, the simple question: How many mutations does it take to evolve a new function?, is usually hard to answer. Here we consider this question in the context of a simple Boolean-network model of a genotype-phenotype map. The nodes and edges of the Boolean network represent the genotype while the Boolean function the network computes is the phenotype. The network evolves by rewiring of its edges, while keeping the number of nodes fixed. We compute the number of mutations needed to achieve a new Boolean function as a function of the number of nodes in the network. We show that the number of mutations scales with the number of nodes of the network as a power-law with an exponent determined by local geometrical features of the genotype-phenotype map. Our theoretical study emphasizes the role of the genotype-phenotype map in determining the time to evolve a new function. 

\end{abstract}

\begin{document}

\flushbottom
\maketitle
\thispagestyle{empty}

\section{Introduction}
A fundamental question of evolutionary dynamics is the time, or number of mutations, needed to evolve a new function \cite{leon_innovation_2018}. Determining this timescale and what sets it can help in predicting and controlling new viral threats and guide directed evolution of proteins. This timescale depends on many factors such as the rate at which mutations occur, population size, the rate of environmental changes, and the genotype-phenotype map \cite{gillespie_role_1999, fragata_evolution_2019}. Here our focus is how the properties of genotype-phenotype map influence the timescale for evolving a new function.

The genotype to phenotype map is defined over the set of all genotypes, with each genotype mapped to a specific phenotype. The set of genotypes can be represented as a genotype graph in which each node represents a genotype and two nodes are connected by an edge if they differ by a single point mutation \cite{wright1932roles}. The genotype graph is typically large and high-dimensional \cite{manrubia_challenge_2026}. For example, the genotype graph for a typical prokaryotic gene consisting of 1000 bases has $4^{1000} \approx 10^{600}$ nodes, representing different 1000 letter genotypes made up of the four nucleotides A, T, C, G. Each node in this genotype graph has 3000 edges, each edge representing a genotype which is reachable by a single point mutation. The large number of possible mutations from each node makes the graph high-dimensional.

The distribution of phenotypes on the genotype graph is crucial in determining the timescale of adaptation because evolution can be thought of as a search process on the genotype graph that starts from an initial genotype and through mutations ends at a genotype that corresponds to the target phenotype \cite{ahnert_structural_2017, de_visser_empirical_2014}. Despite the large and high-dimensional nature of genotype graphs, previous studies have shown that evolution of new functions or higher fitness can be achieved by a surprisingly small number of mutations. Simulations of evolutionary walks consisting of only beneficial mutations on an empirically derived genotype graph for an E. coli protein showed that most walks reached a fitness peak within a handful of mutations \cite{papkou_rugged_2023}. Computational studies on genotype graphs derived for RNA secondary structure and protein sequence models showed most phenotypes can be reached from a given genotype within 15-20 single point mutations \cite{schuster1994sequences, ferrada_comparison_2012}. Other examples include a switch in terpene synthase product specificity with just nine mutations \cite{omaille_quantitative_2008} and a 100,000-fold increase in antibiotic resistance conferred by five mutations in a specific enzyme \cite{weinreich_darwinian_2006}. 

Prior studies have considered features of genotype graphs that can speed up or slow down the time required to evolve a target phenotype such as: i) number of accessible paths to the fitness peak \cite{papkou_rugged_2023, franke_evolutionary_2011}, ii) number of genotypes that represent the target phenotype \cite{schaper_arrival_2014, louis_contingency_2016, cowperthwaite_ascent_2008}, iii) presence of neutral regions within the genotype graph, which contain genotypes with identical phenotypes connected by single point mutations \cite{bakhtin_evolution_2021, rotrattanadumrong_experimental_2022, van_nimwegen_metastable_2000}.

A broader property of the genotype graph that may shape the above features is its dimensionality. In this study, we ask how systematically tuning genotype graph dimensionality affects the number of mutations required to evolve a new function. We define dimensionality as the total number of single mutant neighbors of a given genotype. If the genotype has $N$ components -- these components could be nucleotides in a gene or amino acids in a protein or proteins in a biochemical network -- and each component can be independently drawn from a set of size $C$, ($C = 4$ for nucleotides and $C=20$ for standard amino acids), then the dimensionality of the genotype graph is $N(C-1)$. For a gene containing 1000 bases, the dimensionality is $3000$, whereas for a typical protein with 100 amino acids, the dimensionality is $1900$. Given that biological systems typically have high-dimensional genotype graphs, does the dimensionality make it easier or harder to evolve a function? 

One intuitive way to think about the effect of dimensionality of the genotype graph on the number of mutations to evolve a new function is that if a small number of specific mutations leads to a new function, then finding the right combination among many possible mutations might be difficult leading to a long adaptation time. A recent theoretical study suggests that the time to find a new phenotype can grow exponentially with genotype length \cite{chatterjee_time_2014}. On the other hand, having a large number of potential mutations available to a genotype could open more paths in genotype space via neutral or beneficial mutations, potentially making it easier to reach the target phenotype \cite{franke_evolutionary_2011,cariani_extradimensional_2002, gavrilets_evolution_1997, wu_adaptation_2016}.

Motivated by these studies, we computationally tune the dimensionality of a genotype graph, and ask how that affects the number of mutations needed to evolve a new function. We use the genotype graph associated with Boolean networks. Our study is inspired by and builds upon previous work in \cite{kashtan_spontaneous_2005}. We chose Boolean-network based genotype graph as a model system for several reasons. First, there is an explicit mapping between genotype, phenotype and fitness \cite{kashtan_spontaneous_2005}. The genotype is defined by nodes and edges of the Boolean network, and the phenotype is the Boolean function that the network computes on its inputs. Second, Boolean genotype-phenotype maps have  properties that are analogous to those found in real, biological genotype-phenotype maps such as evolvability and robustness \cite{camargo_boolean_2020, Raman2011}. More generally, Boolean networks have been used to model a number of different cellular networks such as gene regulatory networks, metabolic networks and signal transduction networks \cite{bornholdt_boolean_2008}. Similarly, biochemical networks can be engineered to compute Boolean-like input-output functions \cite{bashor_complex_2019}.

The key observations of our computational study are scaling relations between the number of mutations needed to evolve a target phenotype and the dimensionality of the genotype graph. We explain the observations through a simple analytic model that highlights the features of the genotype graph that determine the scaling exponent.

\section{Results}
\subsection{Computational evolution of a Boolean network}
\begin{figure}[htbp]
  \centering
  \includegraphics[scale =0.79]{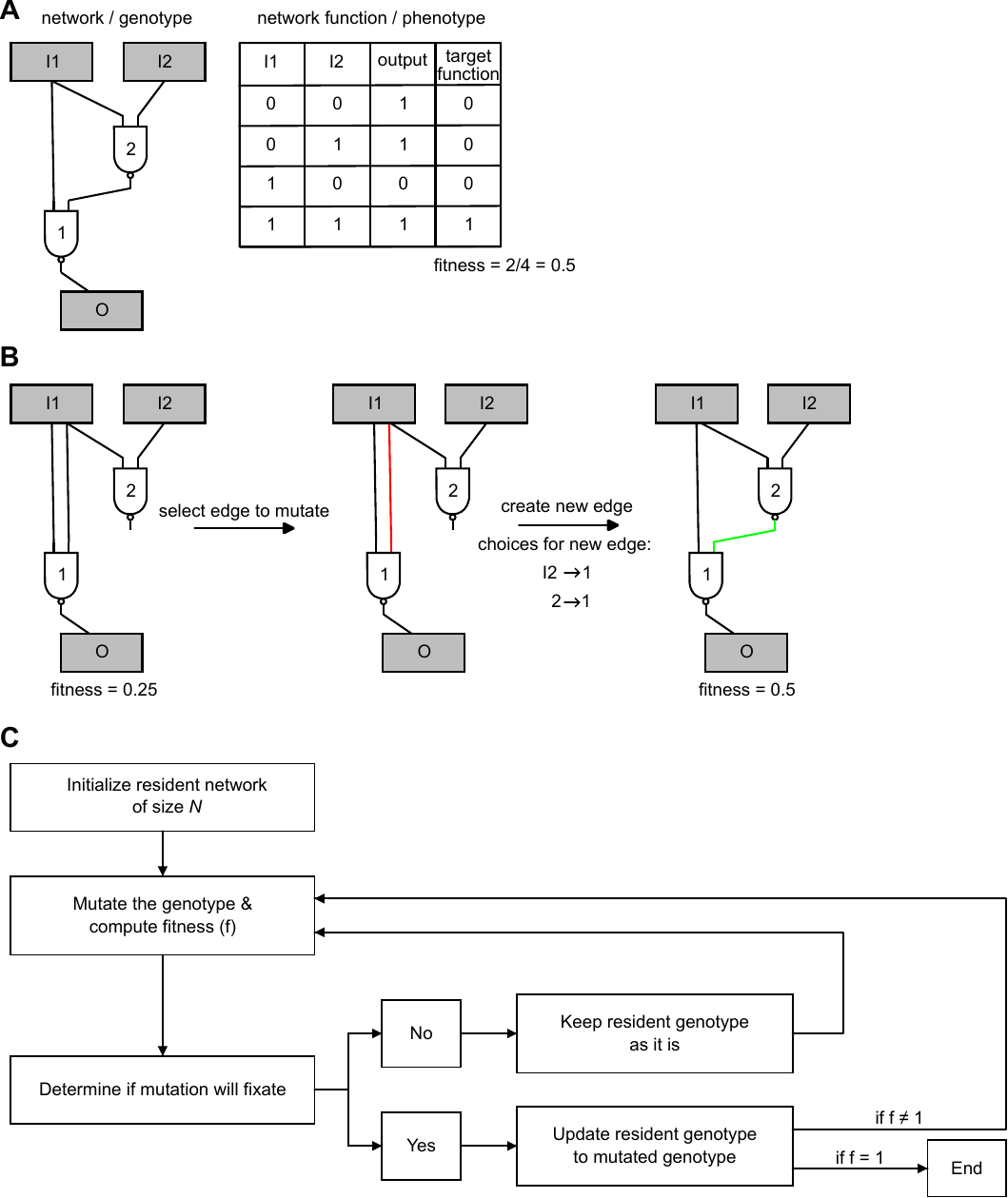}
  \caption{Evolving Boolean networks using a framework adapted from \cite{kashtan_spontaneous_2005}. 
  \textbf{(A)} Example of a Boolean network consisting of two NAND gates (marked as 1 and 2), two external inputs I1 and I2, and an output node O. We define the network size as the number of NAND gates, which in this case is 2. For each external input pair, the network computes an output. The table shows input-output function of the network, which is its phenotype. Fitness is defined as the fraction of network outputs that match the target outputs. 
  \textbf{(B)} An example of a mutation on a network. A mutation occurs on a randomly selected edge (I1 $\rightarrow$ 1, shown in red), which is then removed from the network. This results in NAND gate 1 missing an input edge, which is restored by creating a new edge that connects the output of a randomly chosen node to NAND gate 1 (2 $\rightarrow$ 1, shown in green). This mutation leads to an increase in network fitness.
  \textbf{(C)} Algorithm for evolving Boolean networks. The simulation is initialized with a resident genotype, and a mutation is introduced. The mutation fixates or is lost depending on the fitness of the mutant network, following Moran dynamics. If the mutation fixates, the resident genotype is updated to the mutant genotype. Another mutation is introduced, and the process continues until the target phenotype is achieved (fitness = 1).}
  \label{fig:fig1}
\end{figure}

We define a Boolean network as a directed graph with two external input nodes (I1 and I2), $N$ NAND logic gates, and a single output node O (Fig. \ref{fig:fig1}A; \cite{kashtan_spontaneous_2005}). The size of the network is the total number of NAND gates. Each NAND gate receives two inputs, which can come from the input nodes I1, I2, or as outputs from other NAND gates. The output of a NAND gate is the NAND operation on its two inputs, and the output can serve as an input to any of the other gates.  We define the output of the entire network as the value of the output node O. O is connected to one of the NAND gates and O's value is set by the output of the NAND gate it is connected to. The network as a whole therefore takes two binary values as inputs I1 and I2 and returns a single Boolean value at the output node O.  An example of a Boolean network with two NAND gates, with its corresponding input-output function is shown in Fig. \ref{fig:fig1}A.

The `genotype' of the network is defined by its wiring diagram, which is the list of all connections (edges of the network) between I1, I2, O and the $N$ NAND gates (nodes of the network). The phenotype of the network is the Boolean function it performs on the two external inputs I1 and I2. With two external inputs there are $2^{4}=16$ possible Boolean functions/phenotypes. Since NAND is a universal logic gate, any Boolean function can be constructed using these gates \cite{sheffer_set_1913}. 

To evolve networks from a starting phenotype to a chosen target phenotype, fitness of the network is defined as the fraction of the four external input pairs (I1, I2) for which the network output matches the output corresponding to the target phenotype (\cite{kashtan_spontaneous_2005}). Therefore, the fitness can take five values: 0, 0.25, 0.5, 0.75 and, 1. In Fig. \ref{fig:fig1}A, the output matches for two input pairs, thus the network has a fitness of 0.5.

Next, we define mutations on the Boolean network as rewiring of the connection between nodes, while keeping the network size fixed. More precisely, a rewiring mutation starts by randomly selecting one edge of the network, deleting it, and then replacing it with an output edge from a randomly chosen node (i.e., either one of the $N$ NAND gates, or I1, or I2; see Methods). For example, in Fig. \ref{fig:fig1}B, the edge from I1 to NAND gate 1 is replaced with the output edge from NAND gate 2 to NAND gate 1. This particular mutation changes the fitness of the network from 0.25 to 0.5, assuming that the logic function AND is the target phenotype.  

Mutations that lead to a cycle are not allowed in our simulations. A cycle in a directed network is a path that starts at one node and ends on the same node, while traversing the directed edges of the network. The reason we exclude these networks from consideration is that they do not define an input-output function. As the inputs are propagated through the network from one node to the other via directed edges, cycles lead to a situation where the output is effectively fed back to the input, making the computation ill-defined. In other words, we limit the space of network topologies, or the genotype space, to directed acyclic networks connecting the two external input nodes, the $N$ NAND gates and the output node O. 

We simulate the evolutionary dynamics assuming a haploid Moran process in the strong selection, weak mutation limit. In this limit, selection occurs on a time scale that is much faster than the occurrence of mutations. By considering this regime, we do not need to keep track of each individual network in the population, rather we can represent the population by a single network/genotype. For the haploid Moran process, the probability that a mutant network fixates, i.e., takes over the population, depends on the difference ($s$) between mutant's fitness and the fitness of the original network, as well as the population size ($N_P$) and is given by Eq. \ref{eq1} \cite{mccandlish_epstein_plotkin_2015, sella_hirsh_2005}:
\begin{equation}
P_{\text{fix}} = \frac{1 - e^{-s}}{1 - e^{-N_P s}}
\label{eq1}
\end{equation}

When a neutral mutation occurs (i.e., $s = 0$), then $P_{\text{fix}} = 1/N_P$. 

The simulation starts with an initial network/genotype, which has a fixed number of NAND gates ($N$) and a specified phenotype, i.e., it computes a specific input-output function. A mutation then happens at a random edge of the network leading to a new genotype which competes with the resident genotype. The mutant network either fixates or is lost, with the probability of fixation given by Eq. \ref{eq1}. If the mutation fixates, it becomes the new resident genotype, otherwise the original genotype remains as the resident genotype. This process repeats until the target phenotype is reached, i.e., the fitness is equal to 1 (see Methods). Our simulation scheme is illustrated in Fig. \ref{fig:fig1}C. 

\FloatBarrier

\begin{figure}[htbp]
  \centering
  \includegraphics[scale=0.75]{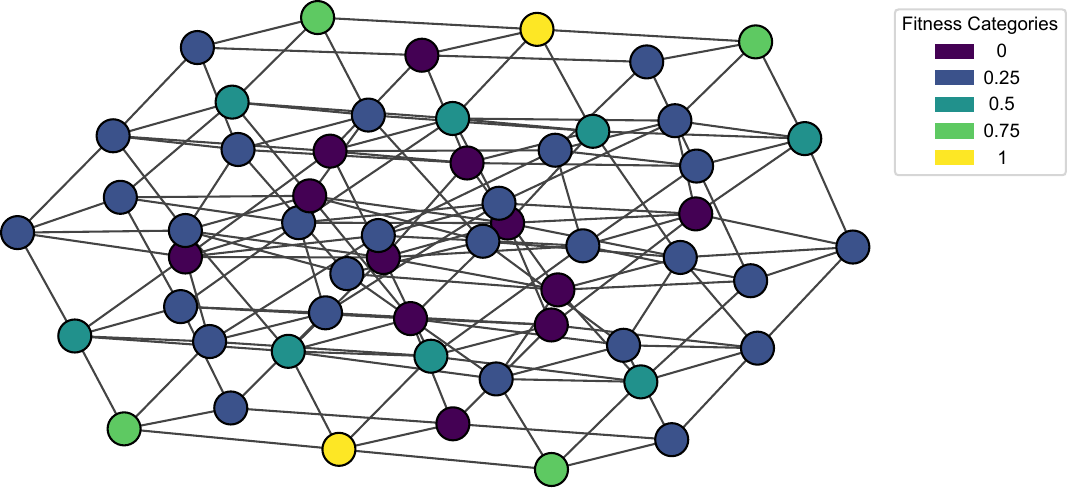}
  \caption{Genotype graph for Boolean networks of size 2. Each node represents an acyclic Boolean network/genotype of size 2, as illustrated in Fig. \ref{fig:fig1}A. Edges connect genotypes that differ by a single mutation. The colors represent the fitness of each network assuming AND as the target phenotype.}
  \label{fig:fig2r}
\end{figure}

The evolution of the Boolean networks produces an adaptive walk on the graph of genotypes, in which nodes correspond to genotypes and edges connect two genotypes related by a single mutation; see Fig. \ref{fig:fig2r}.  Each one of the $N$ NAND gates in the network takes two inputs, each of which can come from one of the $N-1$ other gates, or I1 or I2. This means that the number of networks (genotypes) grows asymptotically as $N^{N}$ for large $N$, whereas there are only a total of 16 distinct logic functions (or phenotypes). In the genotype graph, each network is connected to $O(N^{2})$ neighboring networks, where neighboring networks are connected by a single rewiring mutation (Fig. \ref{fig:fig1}B). The $O(N^{2})$ scaling comes from $2N+1$ choices for which network edge to remove and $O(N)$ choices for the replacement edge. In this context, the key question we address is, how does the number of mutations to reach a target phenotype scale with $N$. 

A simple null model assumes that (a) each of the 16 phenotypes is equally represented among all genotypes and (b) the 16 phenotypes are uniformly distributed on the genotype graph. Under these assumptions, for a given genotype and $N \gg 1$, on average $1/16$ of its $O(N^{2})$ neighboring genotypes will have the target phenotype. A random walk starting from any genotype will reach the target phenotype in an average of 16 mutations. In other words, under this null model, the number of mutations required to reach the target phenotype is independent of $N$.
 
However, genotype graphs can deviate from these assumptions, leading to non-uniform organization of fitness values, as already seen in the genotype graph for $N=2$ networks (Fig. \ref{fig:fig2r}). In Fig. \ref{fig:fig2r}, networks with the NAND phenotype (fitness 0; purple nodes) are primarily neighbors of networks with the same phenotype (fitness 0; purple nodes) or of networks with fitness 0.25 (blue nodes), but rarely neighbor networks with the target phenotype AND (fitness 1; yellow nodes). Consistent with this observation, the number of steps needed to reach a target phenotype through random walk is not independent of network size (Fig. \ref{fig:figS1}). Understanding how the distribution of fitness values in the genotype graph and other properties of genotype--phenotype map affect the number of mutations needed to reach a target phenotype is the key goal of this paper.

\FloatBarrier

\subsection{Number of mutations to evolve AND from NAND scales linearly with network size}
The key question is, how does the number of mutations to reach a target phenotype scale with the size of the network? To address this question we computationally evolve Boolean networks over a large range of network sizes.

\begin{figure}[htbp]
  \centering
  \includegraphics[scale=0.8]{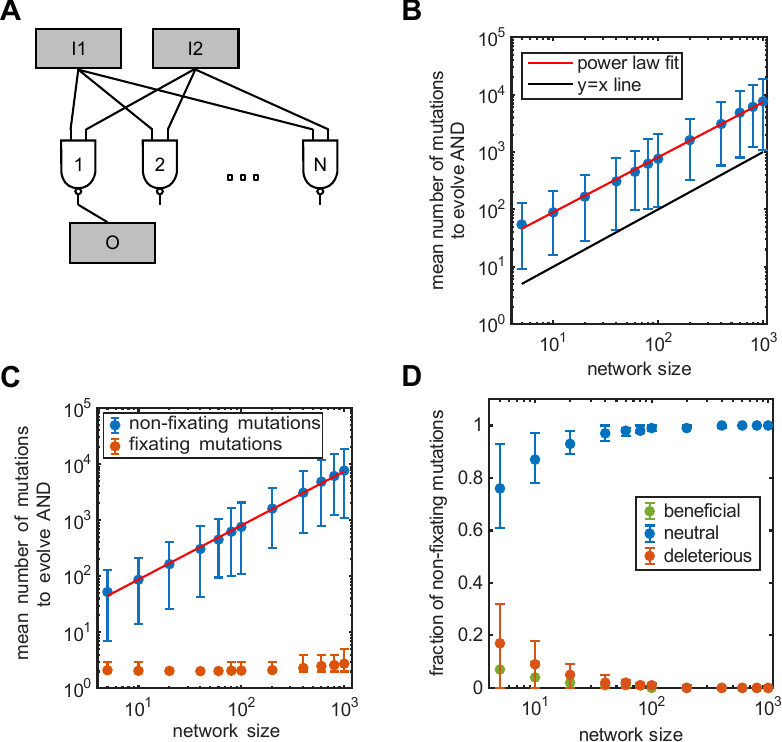}
   \caption{Effect of network size on evolving Boolean network from NAND to target phenotype AND. 
  \textbf{(A)} Initial network configuration for a network of size $N$. The network consists of $N$ NAND gates, two external inputs and one output node O. The inputs for each gate are I1 and I2. This network computes the logic function I1 NAND I2, and its fitness is 0. 
  \textbf{(B--D)} For each network size ($N$), 500 independent simulations were performed, initialized by the network shown in Fig. \ref{fig:fig3}A, using $N_p=10^4$. Dots represent the mean statistic obtained from simulation repeats at each network size and error bars indicate the 5th and 95th percentiles across simulation repeats (see Methods). 
  \textbf{(B)} Mean number of total mutations (including both fixating and non-fixating mutations) required to evolve to a network that computes I1 AND I2. The red line shows a power-law fit, $y=Ax^b$, to all 12 data points, with $A=9.8$ and $b=0.96$. To test the sensitivity of the fit parameters, we also fit three non-overlapping subsets of the data, which gave \(A = 9.6 \pm 3.7\) and \(b = 0.96 \pm 0.11\), reported as mean \(\pm\) standard deviation of fit parameters across the subsets. Black line corresponds to $y=x$ for reference. 
  \textbf{(C)} Mean number of fixating and non-fixating mutations needed to evolve AND as a function of network size. The red line corresponds to $y=9.5x^{0.96}$, the best fit to non-fixating mutations.
  \textbf{(D)} Mean fraction of non-fixating mutations that are beneficial, deleterious or neutral, for each network size.}
  \label{fig:fig3}
\end{figure}

Every simulation is initialized with a genotype of fixed size $N$ that has a NAND phenotype, as shown in Fig. \ref{fig:fig3}A. In this initial network, the inputs for each NAND gate are I1 and I2, and the output node O is connected to NAND gate 1. This is true irrespective of the number of NANDs in the network. Thus, the  simulation starts with fitness 0 and runs until the target phenotype is evolved (fitness 1), using the algorithm described in Fig. \ref{fig:fig1}C. The results in Figs. \ref{fig:fig3}--\ref{fig:fig5} correspond to choosing AND Boolean function as the target phenotype and $N_P$ as $10^4$. In Supplementary, we show results for evolving AND at a smaller population size ($N_P=10$) and for evolving other target logic functions (Figs. \ref{fig:figS3} \& \ref{fig:figS4}A).

We varied the network size from $N=5$ to $1000$, and ran 500 independent simulations for each value of $N$. In each simulation, we first measured the total number of mutations (counting both mutations that fixate and those that do not fixate) needed to evolve to the target phenotype. We count both fixating and non-fixating mutations because they both contribute to the timescale of evolution. Fig. \ref{fig:fig3}B shows that the average number of mutations over the 500 runs needed to reach the target phenotype scales approximately linearly with $N$. The distribution of the number of mutations to reach the target phenotype at a given network size has a tail extending to larger values, though using median instead of the mean as the typical number of mutations to reach the target does not change the scaling (Fig. \ref{fig:figS2}).

To better understand the nature of the adaptive walk from the initial network with fitness zero to the target network with fitness one, we counted the average number of fixating and non-fixating mutations that occurred for each network size during the course of the simulation. Fig. \ref{fig:fig3}C shows that the number of non-fixating mutations grows linearly with the network size while the number of fixating ones is independent of network size. The non-fixating mutations are dominated by neutral mutations, whose fraction approaches one as the network size increases above $N=10$ (Fig. \ref{fig:fig3}D). Fixating mutations are typically beneficial across network sizes. At $N_P=10^4$, the probability of fixating a deleterious mutation is practically zero, while the probability of fixating a neutral mutation is $1/N_P=10^{-4}$. Given that the number of neutral mutations occurring during a simulation ranges from $10^2$ to $10^4$ across network sizes, the expected number of neutral fixations approaches one only at larger network sizes.

Putting it all together, our simulation results (Fig. \ref{fig:fig3}B--D) suggest that the linear scaling of the total number of mutations to reach the target phenotype with network size ($N$) is the result of the linear scaling of the number of neutral non-fixating mutations with $N$. A roughly linear scaling is also observed at $N_P=10$, both for evolving AND and for other target logic functions (see Figs. \ref{fig:figS3} \& \ref{fig:figS4}A).

Next, we identify a geometrical feature of the genotype graph that accounts for the scaling observed. 
\FloatBarrier

\subsection{Geometry of the genotype-phenotype map determines the observed  scaling of number of mutations with network size}
The two main observations from our evolutionary dynamics simulations are that the number of neutral non-fixating mutations needed to evolve AND increases linearly with network size, and that the number of beneficial fixating mutations is independent of network size (Fig. \ref{fig:fig3}C--D). The last observation simply follows from the fact that in the Boolean network model with two external inputs, there are only five possible values of fitness ($f=$ 0, 0.25, 0.5, 0.75, and 1). Evolving AND from NAND requires fitness to increase from 0 to 1 and since each beneficial fixating mutation increases fitness, there will be at most 4 beneficial fixating mutations in the simulation, irrespective of network size.

\begin{figure}[htbp]
  \centering
  \includegraphics[scale=0.74]{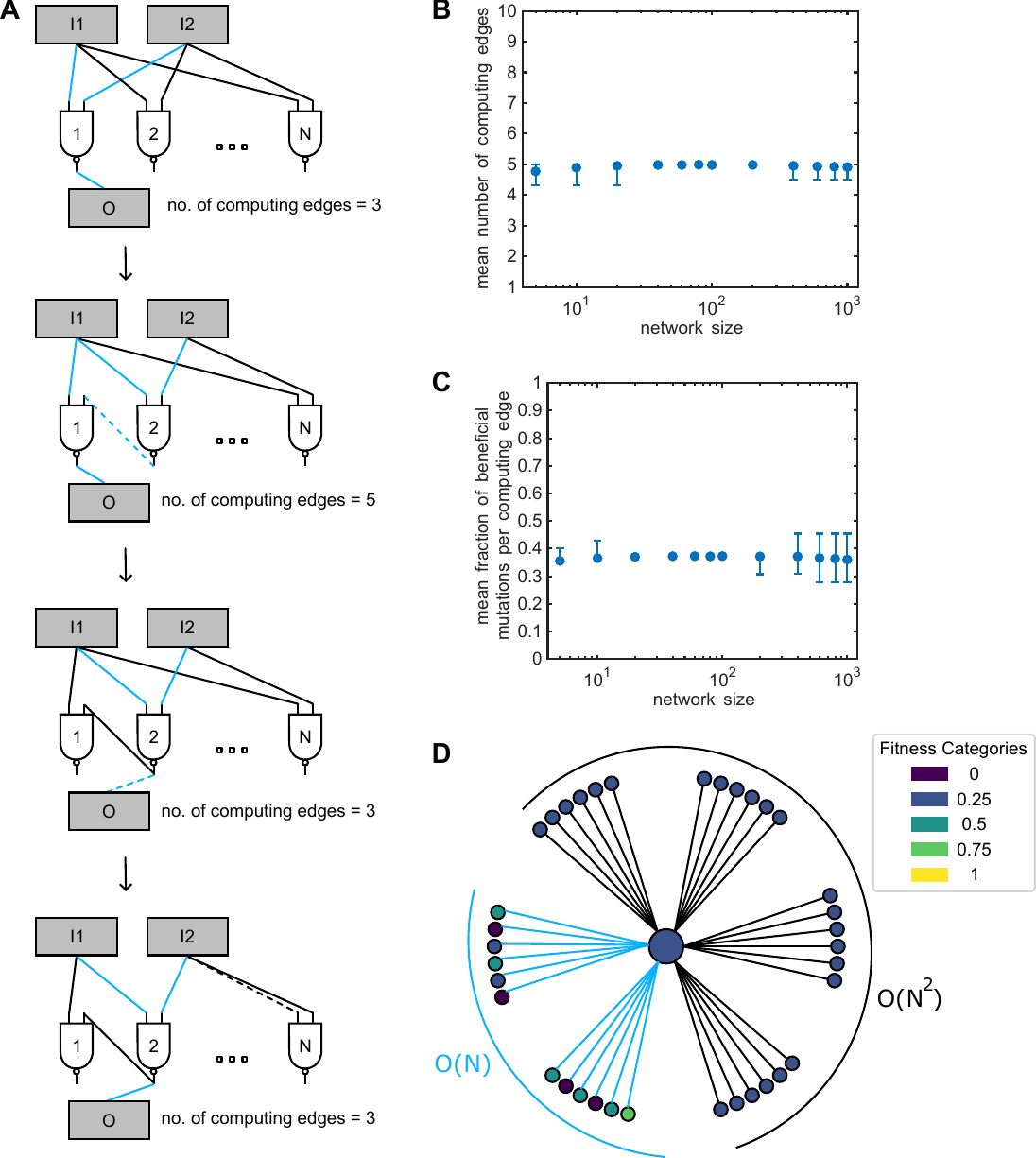}
  \caption{Origin of linear scaling of mutations with network size. 
  \textbf{(A)} A given network has two types of edges: edges that lie on path to the output node O (computing edges, shown in blue) and edges that do not lie on path to O (non-computing edges, shown in black). Arrows indicate a series of sequential mutations to the network, which can change the number of computing edges in the network. In each mutated network, the newly added edge is shown by a dashed line.
  \textbf{(B--C)} Dots represent the mean statistic obtained from simulation repeats and error bars indicate the 5th and 95th percentiles across simulation repeats (see Methods). For some data points, error bars fall within the markers and are therefore not visible. Panels B--C use the same simulation runs as Fig. \ref{fig:fig3}.
  \textbf{(B)} Average number of computing edges in networks during simulations as a function of network size. 
  \textbf{(C)} Fraction of beneficial mutations available per computing edge, as a function of network size.
  \textbf{(D)} Local neighborhood of a genotype. Individual nodes represent different Boolean networks (genotypes) whose fitness is represented by different colors. In this local genotype graph, the node at the center is connected to $O(N^2)$ other nodes by edges that represent single mutations to the Boolean network. The genotypes are grouped into clusters (6 in this case) each one defined by all possible mutations to an edge in the Boolean network at the center of the graph. The mutations to the central Boolean network are either to its non-computing edges (shown in black in panel A) or to computing edges (shown in blue in panel A). In the genotype graph shown here, blue lines connect genotypes related by a mutation to a computing edge, while the black lines represent genotypes connected by mutation to a non-computing edge. For a given computing edge, $P(b|c)$ is the fraction of beneficial mutations among all possible mutations to that edge, as represented by its corresponding cluster. The number of neighboring genotypes connected to the central one by a mutation to a computing edge is $O(N)$.}
  
  \label{fig:fig4}
\end{figure}

Since only a few beneficial fixating mutations are needed to evolve AND, we hypothesized that the observed linear scaling accounts for the number of mutations needed to hit upon a beneficial mutation. To test this hypothesis, we introduce the idea of a computing edge.
An edge in the Boolean network is computing if it lies along a path that connects the external input nodes to the output node O. For example, in Fig. \ref{fig:fig4}A, the edges I1 $\rightarrow$ 1, I2 $\rightarrow$ 1, 1 $\rightarrow$ O are computing edges, while the remaining edges are non-computing. If a computing edge is mutated, the rewired network can have a different fitness, however replacements of non-computing edges always result in a neutral mutation, as non-computing edges are not on a path to the output node O. Mutations can also lead to changes in the number of computing edges. For example, in Fig. \ref{fig:fig4}A, removing the edge I1 $\rightarrow$ 1 and replacing it with 2 $\rightarrow$ 1 increases the number of computing edges from three to five, since I1 $\rightarrow$ 2, I2 $\rightarrow$ 2, I2 $\rightarrow$ 1, 1 $\rightarrow$ O, and the newly established 2 $\rightarrow$ 1 edge will all be computing. An additional mutation that replaces the 1 $\rightarrow$ O edge with the 2 $\rightarrow$ O edge reduces the number of computing edges back to three, as these two mutations effectively swap NAND gates 1 and 2.  

To encounter a beneficial fixating mutation, the mutation must occur on a computing edge and lead to higher fitness. Therefore, the probability of making a mutation that is beneficial and fixates is:

\begin{equation}
P(b) = P(c) \cdot P(b|c) \cdot P_{\text{fix}}
\label{eq2}
\end{equation}

where $P(c)$ is the probability of choosing a computing edge to mutate, and is equal to the fraction of network edges that are computing. $P(b|c)$ is the conditional probability that a rewiring mutation increases fitness, given that a computing edge was chosen (see Fig. \ref{fig:fig1}B for the mutation process); $P_{\text{fix}}$ is the probability of fixating that mutation.

For the simulations shown in Fig. \ref{fig:fig3}, we find that the average number of computing edges in networks generated during evolution is independent of network size (Fig. \ref{fig:fig4}B). This is expected because the number of computing edges in the starting network configuration is three, independent of network size, and the adaptive walk leading to the target phenotype (AND) contains only few fixating mutations at population size $N_p= 10^4$. Since the network has a total of $2N + 1$ edges, the probability of choosing a computing edge to be mutated, $P(c)$, is on the order of $1/N$. We measured $P(b|c)$ from the simulations by counting the fraction of beneficial mutations available among all possible mutations to a computing edge, and observed that $P(b|c)$ is independent of network size (Fig. \ref{fig:fig4}C). $P_{fix}$ is computed from the Moran process, and is also independent of network size. Therefore, the probability of obtaining a beneficial fixating mutation scales as $P(b) \sim 1/N$. Finally, we conclude that the average number of mutations that occur before a beneficial fixating one is randomly chosen is $1/P(b) \sim N$, which explains the scaling observed in Fig. \ref{fig:fig3}C. Since practically all these mutations will occur on non-computing edges, they will be neutral and will rarely fixate given the large population size. This explains the observations that non-fixating mutations scale linearly with $N$ (Fig. \ref{fig:fig3}C), and that most non-fixating mutations are neutral (Fig. \ref{fig:fig3}D). 

In Fig. \ref{fig:fig4}D, we explain the scaling of $P(b)$  with Boolean network size (Fig. \ref{fig:fig3}B--D) from the perspective of the local geometry of the genotype graph. By local geometry, we mean the organization of phenotypes in the local single-mutation neighborhood of a genotype. Evolution can be thought of as a walk on the genotype graph that starts at a genotype that computes NAND and through fixating mutations ends at a genotype that computes AND. A typical genotype (central node in the genotype graph in Fig. \ref{fig:fig4}D) is connected to $O(N^2)$ other genotypes via single edge mutations. Of the $O(N^2)$ genotypes, $O(N)$ are obtained through mutation to a computing edge in the central genotype. This follows from the fact that for a given genotype, the number of computing edges is of order one, and for each computing edge, there are on the order $O(N)$ possible rewiring mutations. If the fraction of these $O(N)$ genotypes that have fitness higher than the central genotype is fixed and independent of $N$ (see Fig. \ref{fig:fig4}C), then the probability that a randomly chosen mutation is beneficial among all $O(N^2)$ single-mutation neighboring genotypes is of order $N/N^2 = 1/N$. 

So far we have considered a population size $N_p=10^4$. At a much smaller population size, $N_p=10$, the total number of mutations (including fixating and non-fixating mutations) required to evolve AND still scales roughly linearly with network size (Fig. \ref{fig:figS4}A; scaling exponent $=0.9$). We show that $P(c)P(b|c)$ scales approximately as $N^{-0.87}$ (Fig. \ref{fig:figS4}C--D), implying that on average the number of mutations needed to encounter a beneficial mutation grows roughly linearly with network size. While at $N_p=10^4$ these $O(N)$ mutations are predominantly non-fixating, at $N_p=10$ they include a mix of fixating and non-fixating mutations, owing to the increased probability of fixating neutral mutations. Consequently, the resident network is replaced by a new network configuration more often during the evolutionary walk (Fig. \ref{fig:figS4}B). Despite more frequent changes to the network configuration, the observed scaling suggests that the geometric picture illustrated in Fig. \ref{fig:fig4}D remains applicable during evolution.

\FloatBarrier

\subsection{Effect of perturbing the local geometry of the genotype to phenotype map on the number of mutations to evolve AND}
\begin{figure}[htbp]
  \centering
  \includegraphics[scale=0.78]{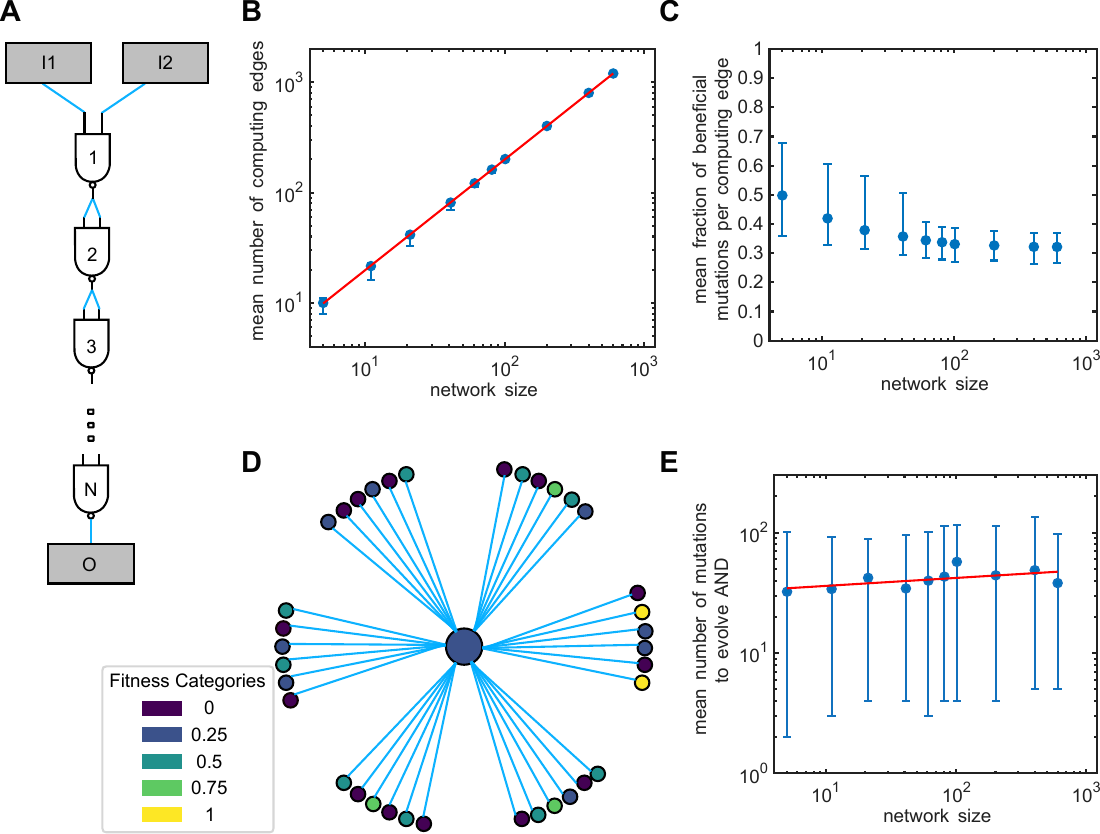}
  \caption{Changing the initial network configuration changes scaling of number of mutations with network size for evolving target phenotype AND. Results shown here are for $N_p=10^4$. \textbf{(A)} Network configuration for initializing the evolution simulation. In this network, all edges are computing edges. 
  \textbf{(B--D)} Analogous to Fig. \ref{fig:fig4}B--D, but with evolution initiated from the network in Fig. \ref{fig:fig5}A. In panels B--C, error bars represent the 5th--95th percentiles across simulation repeats and are not visible for some data points because they fall within the marker.
  \textbf{(B)} Mean number of computing edges in the network during the simulation, as a function of network size. The red line represents $y=2x^{1}$, the best fit to the data. Fitting two non-overlapping halves of the data also gave scaling exponents of $0.99$ and $1$.
  \textbf{(C)} Mean fraction of choosing a beneficial mutation given a computing edge, as a function of network size. 
  \textbf{(D)} Schematic of the local neighborhood of the genotype corresponding to the Boolean network in panel A. All mutations to the central Boolean network are to its computing edges, unlike Fig. \ref{fig:fig4}D. The number of neighboring genotypes connected to the central one by a mutation to a computing edge is therefore $O(N^2)$.
  \textbf{(E)} Analogous to Fig. \ref{fig:fig3}B, but with evolution initiated from the network in Fig. \ref{fig:fig5}A. Dots represent the total number of mutations to evolve AND at a given network size, averaged over simulation repeats. The total number of mutations includes both fixating and non-fixating mutations that occur during the simulation. Error bars indicate the 5th and 95th percentile range across simulation repeats. The red line represents power law fit to $y = 31x^{0.07}$. Fitting two non-overlapping halves of the data gave scaling exponents of $0.07$ and $-0.09$.}
  \label{fig:fig5} 
\end{figure}

To further test whether the two properties--the fraction of computing edges in the network and the probability that mutating a computing edge produces a beneficial mutation--can capture the scaling of number of mutations with network size, we perturbed one of these two properties. 
Specifically, we changed the fraction of computing edges in the Boolean network, by choosing the initial genotype to be the one shown in Fig. \ref{fig:fig5}A. This choice allows us to sample a different local geometry of the genotype graph at the start of the evolutionary walk. We then tested whether the scaling of the number of mutations required to evolve AND with network size $N$ changed as predicted. The results below correspond to simulations at population size $N_p = 10^4$, though the conclusions remain same for $N_p=10$ (see supplementary fig \ref{fig:figS5}). 

In the Boolean network shown in Fig. \ref{fig:fig5}A every edge is computing. This implies that all the mutations we can make to this initial network will hit a computing edge ($P(c) = 1$). As the network evolves to the target phenotype, the average number of computing edges remains of order $N$ (Fig. \ref{fig:fig5}B).  Since the total number of edges in the network is $2N+1$, $P(c)$ remains independent of network size. The mean fraction of beneficial mutations obtained per computing edge mutation decreases for small $N$, but for $N>20$, it remains approximately independent of $N$, similar to the previous case; compare Fig. \ref{fig:fig4}C and Fig. \ref{fig:fig5}C.

Geometrically, this initial genotype confines the adaptive walk to a region of genotype space where a typical genotype's $O(N^2)$ single-mutation neighbors arise from mutations to computing edges. This contrasts with the initial Boolean network chosen earlier (Fig. \ref{fig:fig3}A), where a typical genotype had $O(N^2)$ single-mutation neighbors, but only $O(N)$ of these were accessible via computing-edge mutations. As measured by the simulations, a fixed fraction of these neighbors are beneficial. These two properties of the genotype graph are represented in Fig. \ref{fig:fig5}D. 

Based on the geometry of the genotype graph shown in Fig. \ref{fig:fig5}D and using equation 2, we expect that the probability of obtaining a beneficial fixating mutation to be independent of network size $N$. Therefore, the expected number of total mutations to evolve AND is predicted to be independent of the size of the Boolean network. This prediction is borne out by simulations, as seen in Fig. \ref{fig:fig5}E.

\FloatBarrier

\section{Discussion}
Genotype graphs are typically high dimensional, where dimensionality is defined as the number of genotypes accessible by a single mutation from a given genotype. Some research suggests that the high dimensional nature of genotype graphs can make it easier to evolve a new phenotype by providing many pathways to the fitness peak \cite{franke_evolutionary_2011,cariani_extradimensional_2002, gavrilets_evolution_1997, wu_adaptation_2016}, while other studies suggest that the adaptation time can grow exponentially with genotype length \cite{chatterjee_time_2014,heredia_selection_2017}. Motivated by these studies, here we ask how tuning the dimensionality of a genotype graph affects the number of mutations required to evolve a target phenotype.

To address this question, we simulated evolutionary dynamics towards a target phenotype on genotype graphs of different dimensionalities. We employed Boolean network-based genotype graph and measured the number of mutations to reach a target phenotype as a function of the size of the Boolean network. We define Boolean network size as the number of logic gates in the network ($N$). Increasing $N$ increases the dimensionality of the genotype graph, which scales as $N^2$.

We simulate evolutionary dynamics that start at a specific, initial genotype that performs the NAND logic function and end when the genotype has evolved to perform the target logic function AND. Evolution proceeds via mutations of the network edges, while keeping the network size $N$ fixed (Fig. \ref{fig:fig1}C). We then measure the average number of mutations needed to evolve the target logic function, as a function of $N$. We observe that the total number of mutations (counting both non-fixating and fixating mutations) scales linearly with respect to $N$, at large population size, where neutral mutations rarely fix (Fig. \ref{fig:fig3}B). A roughly linear scaling is also observed at small population size, where neutral mutations fix more frequently (Fig. \ref{fig:figS4}A).

In our simulation setup, evolving the target function AND requires at most four fixating beneficial mutations, because fitness can take only five discrete values. To understand the observed scaling, we examine the probability of obtaining a beneficial mutation. We split this probability into two factors: $P(c)$ which is the fraction of network edges that compute the logic function (``computing edges'') and $P(b|c)$ which is the fraction of fitness-increasing mutations available per computing edge. The average fraction of computing edges scales as $1/N$ with network size (Fig. \ref{fig:fig4}B), while the mean fraction of beneficial mutations per computing edge is constant with respect to $N$ (Fig. \ref{fig:fig4}C). Encountering a beneficial mutation during the evolution requires that a mutation occur on a computing edge and lead to increased fitness. Therefore, the probability of obtaining a beneficial mutation is $P(c)P(b|c)$, which scales as $1/N$. Consequently, the average number of mutations needed before obtaining a beneficial mutation scales as $N$. These two factors are sufficient to explain the observed linear scaling at large population size and also the approximately linear scaling at small population size. 

The mean fraction of computing edges within the network during evolution is set by the initial network configuration. This leads to a simple prediction: changing the local geometry of the initial network can change the scaling of total number of mutations with network size. To test this prediction, we constructed a different initial network in which the fraction of computing edges is independent of $N$ (Fig. \ref{fig:fig5}A). As predicted, the number of mutations to reach AND was roughly independent of network size (Fig. \ref{fig:fig5}E). This scaling was explained by the scaling of $P(c)$ and $P(b|c)$ for both small and large population sizes.

An interesting consequence of these scaling relationships is on the mean time taken to evolve AND. If the occurrence of mutations is described by a Poisson process with mean time $\tau$ per edge, then the total time taken to evolve AND, $T$, is equal to the sum of waiting times between consecutive mutations. Each network has $2N+1$ edges, so for a mutation rate $1/\tau$ per edge, mutations occur across a population of size $N_P$ at rate $\frac{N_P(2N+1)}{\tau}$. The mean waiting time between mutations therefore scales as $1/N$, giving $\langle T \rangle \sim \langle M \rangle/N$, where $\langle M \rangle$ is the average number of mutations to reach the target phenotype. If $\langle M \rangle$ scales linearly with $N$ (Fig. \ref{fig:fig3}B), then the average time to evolve AND will be constant with network size (Fig. \ref{fig:figS6}A), whereas if $\langle M \rangle$ is constant with respect to $N$ (Fig. \ref{fig:fig5}E), larger sized networks will evolve the target function faster compared to smaller networks (Fig. \ref{fig:figS6}B).

To study how genotype size affects the evolutionary dynamics, we chose initial genotypes such that they were consistent across different genotype sizes, allowing us to isolate the effect of size of genotype. For example, for the initial configuration shown in Fig. \ref{fig:fig3}A, inputs $I_1$ and $I_2$ feed into each NAND gate, irrespective of network sizes. This choice of the set of initial network configurations may set the scaling of $P(c)$ and $P(b|c)$ with network size. The influence of these initial configuration may be strongest when only a few mutations fix. In this regime, the initial network sets the scaling of $P(c)$ and $P(b|c)$ with network size, and a small number of fixating mutations does not substantially alter these properties during the evolutionary walk.

The results discussed here are limited to the extent that they depend on the specific properties of Boolean genotype--phenotype map, that are distinct from those of other computational and experimentally studied maps. Unlike other biological genotype--phenotype maps that can exhibit a range of fitness values, the Boolean genotype-phenotype map considered here has exactly five fitness levels. This limits the number of beneficial fixating mutations needed to reach a desired function to less than five. The number of phenotypes is also limited and independent of the network size in the Boolean network. Five NAND gates are sufficient to encode all sixteen phenotypes corresponding to all possible Boolean functions of two external inputs. In contrast, for biological genotype--phenotype maps, larger genotypes could encode more phenotypes.

While how our specific scaling results generalize to other choices of initial conditions and genotype--phenotype maps is unclear, our results show that the dimensionality of genotype space alone does not determine how difficult it is to evolve a new function; rather, a relevant quantity is the fraction of mutational directions that increase fitness. This fraction can be decomposed as fraction of directions capable of changing the phenotype ($P(c)$) and, among those, the fraction of directions that increase fitness ($P(b|c)$). A related concept appears in \cite{greenbury_organization_2015}, which showed that organization of genotype into constrained positions (where mutations can change phenotype; analogous to computing edges in our study) and unconstrained positions (where mutations do not change phenotype) drives fundamental properties of genotype-phenotype maps such as phenotype bias and evolvability. The simulations in this study provide a concrete example of how this organization of the genotype can also shape the timescale of evolutionary search. 

The probability of encountering a beneficial mutation, $P(c)P(b|c)$, may be useful in predicting evolutionary timescales in regimes where adaptation proceeds through a relatively small number of beneficial mutations that are accessible in the local neighborhood of a genotype. Experimental studies show that adaptation can sometimes occur through a handful of beneficial mutations, including in the evolution of drug resistance \cite{weinreich_darwinian_2006, lozovsky_stepwise_2009, papkou_rugged_2023, palmer_delayed_2015} and recovery of motility in immotile bacteria \cite{horton_transient_2023}. Computational analyses of genotype graphs also show that most phenotypes are found from a given genotype through a small number of single point mutations \cite{greenbury_organization_2015, greenbury_tractable_2014, schuster1994sequences, ferrada_comparison_2012}. For genotype graphs of RNA or protein sequences, this number is estimated to be around 15--20 mutations \cite{schuster1994sequences, ferrada_comparison_2012}. Additionally, theoretical studies suggest that the probability that at least one fitness increasing path exists approaches 1 as dimensionality increases \cite{franke_evolutionary_2011, krug_evolutionary_2024, hegarty_existence_2014}. However, the idea embodied by Eq. \ref{eq2} may be insufficient to explain evolutionary dynamics in regimes where no beneficial mutations are immediately accessible in the local neighborhood of a genotype. In such regimes, the time required to traverse neutral regions may become rate-limiting \cite{huynen_smoothness_1996, van_nimwegen_metastable_2000, rotrattanadumrong_experimental_2022}. Determining when the factors described here control evolutionary timescales, and when other features of genotype space set the timescale, remains an open question.

\section{Competing interests}
We declare no competing interests.
\section{Acknowledgments and Funding}
We thank Pankaj Mehta, members of Mehta group, and members of Kondev group for discussions and feedback. This work was supported by grants to JK from the Simons Foundation (MPS-SIP00687713), and The Chan Zuckerberg Initiative. NF was also supported by NIH R01 GM081648. EL was supported by NSF grant DMS-2527337 and Dartmouth College Burke Research Initiation Award. 
\section{Methods and Materials}
\subsection{Algorithm for simulating evolutionary dynamics} 

The goal of the evolutionary simulation is to evolve an initial network configuration that performs a NAND operation between two external inputs I1 and I2 to instead perform a different logic operation (such as AND), similar to \cite{kashtan_spontaneous_2005}. The evolution occurs through mutations on the edges of the network, while keeping the nodes of the network fixed. At any given time point in the simulation, the population is represented by a resident network that competes with a mutant network. 

Each simulation is initialized by the following parameters: population size ($N_P = 10^4$ or $N_P = 10$), mutation rate per edge per network ($1/\tau = 0.001$ per unit time; kept fixed for all simulations), network size ($N$), initial network configuration (Fig. \ref{fig:fig3}A or Fig. \ref{fig:fig5}A) and the target logic function. For simulations initialized with the network in Fig. \ref{fig:fig5}, only odd values of $N$ were used to allow for NAND phenotype at the start of evolution. A given network configuration is stored as a list of directed edges and nodes using NetworkX module in Python. The target logic function is represented in the script as an ordered binary list of four values corresponding to the outputs for input pairs (I1, I2) = (0, 0), (0, 1), (1, 0), and (1, 1). For example, the target logic function AND is stored as [0, 0, 0, 1]. Because the initial network performs NAND function (corresponding to [1, 1, 1, 0]), none of the outputs of the initial network matches the target function AND, thus the initial network has a fitness of 0.  Time is also set to 0 at the start, with the resident network being the initial network.

A mutation occurs after a waiting time sampled from an exponential distribution with rate $\frac{(2N + 1) \times N_P}{\tau}$. The mutation is made on an edge of the resident network (Section \ref{sec:mutation}) and fitness is evaluated for the mutant network (Section \ref{sec:fit}). The probability that the mutant network fixates is given by Eq. \ref{eq1}, which depends on the population size and the fitness difference between the mutant network and the resident network. We then draw a random number from a uniform distribution between 0 and 1 to determine if the mutant network will fixate or not. Irrespective of whether the mutant network fixates or not, we record the timepoint at which the mutation occurred, the type of mutation (beneficial, neutral or deleterious, determined by comparing the fitness change between mutant and resident network), the fate of the mutant network (fixed or not), and the number of computing edges in the mutant network. If the mutant fixates, it becomes the new resident network, and otherwise the resident network stays as it is. The process of generating a mutant network is repeated until the fitness of the resident network reaches 1.

For simulations performed at \(N_P=10\), some simulation runs did not reach the target fitness within the allocated computational time. These incomplete simulations were excluded from the analysis. The exact percentages of excluded simulations for each network size are provided in Supplementary Table S1.

\subsection{Algorithm for mutating a network}
\label{sec:mutation}
To perform a mutation on a network, an edge is randomly selected from all edges in the network. For the selected edge $(u \rightarrow v)$, where $u$ is the source node and $v$ is the target node, we generate a list of replacement options for $u$. If $v$ is the output node O, then replacement options for $u$ are NAND gates except $u$. If $v$ is a NAND gate, then replacement options for $u$ are all nodes except $u$, $v$, and O. A new source node $u'$ is randomly selected from the options, and the original edge $(u \rightarrow v)$ is replaced with the new edge $(u' \rightarrow v)$. The mutated network is then tested for cycles using the \texttt{is\_directed\_acyclic\_graph()} function from the NetworkX library. If the network is acyclic, the proposed mutation is accepted and it either fixates or does not fixate according to Eq. \ref{eq1}. If a cycle is detected, the mutation is rejected as invalid, and the mutation process is restarted by randomly selecting an edge to be mutated. This process is repeated until an acyclic network is obtained.

\subsection{Algorithm for obtaining computing module for a network}
\label{sec:comp}
To obtain the sub-network of a given network that computes the output from the external inputs, we trace back the edges from the output node O to the external input nodes (I1 and/or I2). This sub-network is referred to as the computing module and includes only the nodes and edges that are involved in computing the network's output.

To trace the computing module, we begin at the output node O and identify the NAND gate directly connected to it. Both O and the NAND gate connected to O are added to the computing module, along with the edge between them. The algorithm then iteratively works backward: for each NAND gate that was just added to the computing module, we identify its two input nodes and add both the input nodes and their connecting edges to the computing module. This process continues until all newly added nodes are external inputs (I1 and/or I2), at which point the backward traversal terminates.

\subsection{Algorithm for evaluating network fitness}
\label{sec:fit}

We use the definition of fitness, as given in \cite{kashtan_spontaneous_2005}. Fitness of the network is the fraction of external input combinations for which the network output matches the target logic function.

To evaluate the fitness of a mutated network, we first identify its computing module (Section \ref{sec:comp}). Each mutation replaces an edge $(u \rightarrow v)$ in the resident network with a new edge $(u' \rightarrow v)$. If $v$ is not part of the computing module, then the mutation occurs outside the part of the network that determines the output, and the mutant network is assigned the same fitness as the resident network. If $v$ is part of the computing module, the mutation can change the network output, and we determine fitness by computing the mutant network's output for each of the four external input combinations (I1,I2) $\in$ (0,0), (0,1), (1,0), (1,1). This avoids recomputing fitness for mutations that cannot affect the network output, thereby speeding up the simulation.

To measure fitness, we propagate I1 and I2 values through the computing module for each pair of external inputs. Each NAND gate is assigned a value as the NAND operation between the values of its input nodes. Gates are assigned values iteratively: a gate is assigned a value once both of its inputs have values; otherwise, it waits for the next iteration. This process continues until the gate connected to output node O is assigned a value, which is recorded as the network's output for that external input combination. This process is repeated for all external input combinations.

We then determine the fraction of external input combinations for which the network's output matches the output corresponding to the target logic function. For example, a network that correctly computes 3 out of 4 external input combinations has a fitness of 0.75.

\subsection{Algorithm for measuring fraction of beneficial mutation per computing edge}
\label{sec:fb}
We measure the fraction of beneficial mutations per computing edge for the initial network configuration and for each subsequent network that fixates during the simulation. For each computing edge in the network, represented by $(u \rightarrow v)$ where $u$ is the source node and $v$ is the target node, we generate all possible alternative source nodes for that edge. If $v$ is the output node O, then the alternative source nodes are all NAND gates excluding $u$. If $v$ is a NAND gate, then the alternative source nodes can be all network nodes excluding $u$, $v$, and O.

For each alternative source node $u'$, we temporarily replace the original edge $(u \rightarrow v)$ with the new edge $(u' \rightarrow v)$. If the modified network remains acyclic, we evaluate its fitness and classify the mutation as beneficial (fitness increase), neutral (no fitness change), or deleterious (fitness decrease) relative to the original network. We then revert the network to its original state and repeat the process with a different choice of source node. The fraction of beneficial mutations ($f_b$) for a given computing edge is calculated as the number of beneficial mutations divided by the total number of allowed mutations available for that computing edge.

For a subset of simulations using $N_p=10$, we used an optimized, conceptually equivalent implementation of the above algorithm to reduce computational time. Both algorithms are provided in the data repository.

\subsection{Statistics from evolutionary simulations}
We use the evolutionary simulations described above to measure different quantities of interest, such as the total number of mutations needed to evolve the target logic function. For a given network size, population size, target logic function, and initial network configuration, we repeat the simulation about 500 times independently. In the figures, we report summary statistics for each quantity across these simulation repeats. This section describes how the summary statistics were computed for the following quantities:
\\[0.5em]
\noindent (i) number of fixating mutations, number of non-fixating mutations, total number of mutations (sum of number of fixating and non-fixating mutations), fraction of beneficial, deleterious, and neutral non-fixating mutations, and time to evolve.

\noindent (ii) number of computing edges in a network.

\noindent (iii) fraction of beneficial mutations per computing edge ($f_b$).
\\[0.5em]
For each quantity listed in (i), we record one value per simulation. From the independent simulation repeats, we obtain a distribution of values. The figures report the mean across simulations, with error bars showing the 5th and 95th percentiles of this distribution (Figs. \ref{fig:fig3}B--D, \ref{fig:fig5}E, \ref{fig:figS3}A--C, \ref{fig:figS4}A--B, \ref{fig:figS5}A--B, and \ref{fig:figS6}A--B).

For (ii), we record the number of computing edges for each resident network during a simulation. The resident networks include the initial network and every subsequently fixated network. We then take average of the number of computing edges across these networks, giving one value for each simulation. From the independent simulation repeats, we obtain a distribution of these simulation-level averages. The figures report mean of the distribution, with error bars showing the 5th--95th percentile range (Figs. \ref{fig:fig4}B and \ref{fig:fig5}B). The probability of choosing a computing edge, $P(c)$, is obtained by dividing the mean number of computing edges by the total number of edges in the network, $2N+1$ (Fig. \ref{fig:figS4}C \& \ref{fig:figS5}C).

For (iii), we record $f_b$ for each computing edge of each resident network during the simulation. We then average these $f_b$ values to obtain one mean $f_b$ for each simulation. The figures report the mean of these per-simulation $f_b$ values, with error bars showing the 5th--95th percentile range across independent simulation repeats (Figs. \ref{fig:fig4}C, \ref{fig:fig5}C, \ref{fig:figS4}D, and \ref{fig:figS5}D).

\printbibliography
\clearpage
\begin{center}
\Large\textbf{Supplementary figures for:}\\[0.5em]
\large\textbf{How many mutations does it take to evolve a new function?}\\[0.5em]
\normalsize Nida Farheen, Zachary Boroda, Ethan Levien, and Jan\'e Kondev
\end{center}
\vspace{1em}

\renewcommand{\thefigure}{S\arabic{figure}}
\setcounter{figure}{0}

\begin{figure}[htbp]
  \centering
  \includegraphics[scale=0.85]{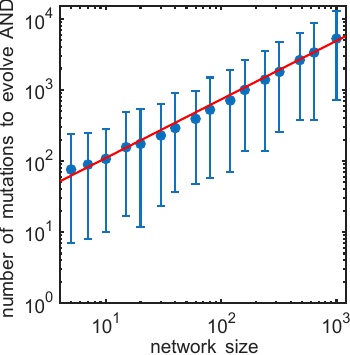}
\caption{Number of mutations required to reach AND phenotype through random walk as a function of network size. In the random-walk simulation, each mutation is accepted regardless of whether it is neutral, beneficial, or deleterious. For each network size, 500 independent simulations were performed, initialized with the network configuration shown in Fig. ~\ref{fig:fig3}A. Dots show the mean number of mutations required to reach AND, and error bars indicate the 5th--95th percentile range across simulation repeats. Red line shows best fit to $y=Ax^b$, where $A=18.7$ and $b = 0.8$. Fitting three non-overlapping subsets of the data produced $b=0.8\pm0.2$ (mean $\pm$ standard deviation of the fitted exponent across the subsets).}
  \label{fig:figS1}
\end{figure}
\FloatBarrier

\begin{figure}[htbp]
  \centering
\includegraphics[scale=0.83]{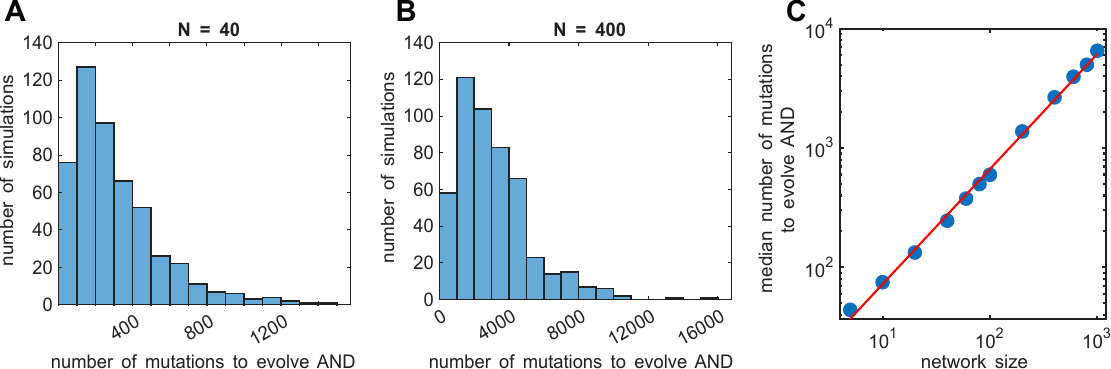}
  \caption{Distribution of total number of mutations required to evolve AND, constructed from 500 independent simulations at a given network size. Simulations are initialized using the network configuration shown in fig. \ref{fig:fig3}A and use $N_p= 10^4$. \textbf{(A)} Mutation distribution for network size $N = 40$. \textbf{(B)} Mutation distribution for $N = 400$. \textbf{(C)} Analogous to fig. \ref{fig:fig3}B, but plotting median rather than the mean number of mutations as a function of network size. The red line shows power law fit, $y = 7.8x^{0.96}$.}
  \label{fig:figS2}
\end{figure}
\FloatBarrier

\begin{figure}[htbp]
  \centering
  \includegraphics[scale= 0.85]{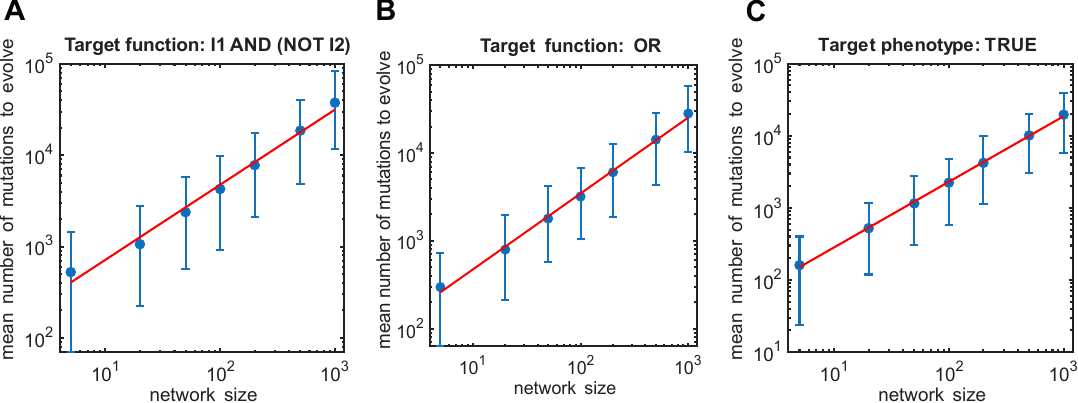}
  \caption{Mean number of mutations required to evolve different logic functions from NAND, using population size $N_P = 10$. Simulations are initialized using the network configuration shown in fig. \ref{fig:fig3}A. Error bars represent the 5th–95th percentile range across simulation repeats and the red line represents a power-law fit $y=Ax^b$ to all data points. Brackets denote the scaling exponents obtained from fitting the first three and last four data points, respectively. \textbf{(A)} target phenotype I1 AND (NOT I2), fitted to $y = 107.8x^{0.82}$ [$b$: $0.64$, $0.95$]. \textbf{(B)} target phenotype OR, fitted to $y = 64.6x^{0.86}$ [$b$: $0.78$, $0.94$]. \textbf{(C)} target phenotype TRUE, fitted to $y=34.8x^{0.91}$ [$b$: $0.86$, $0.94$].}
  \label{fig:figS3}
\end{figure}
\FloatBarrier

\begin{figure}[htbp]
  \centering
  \includegraphics[width=0.8\textwidth]{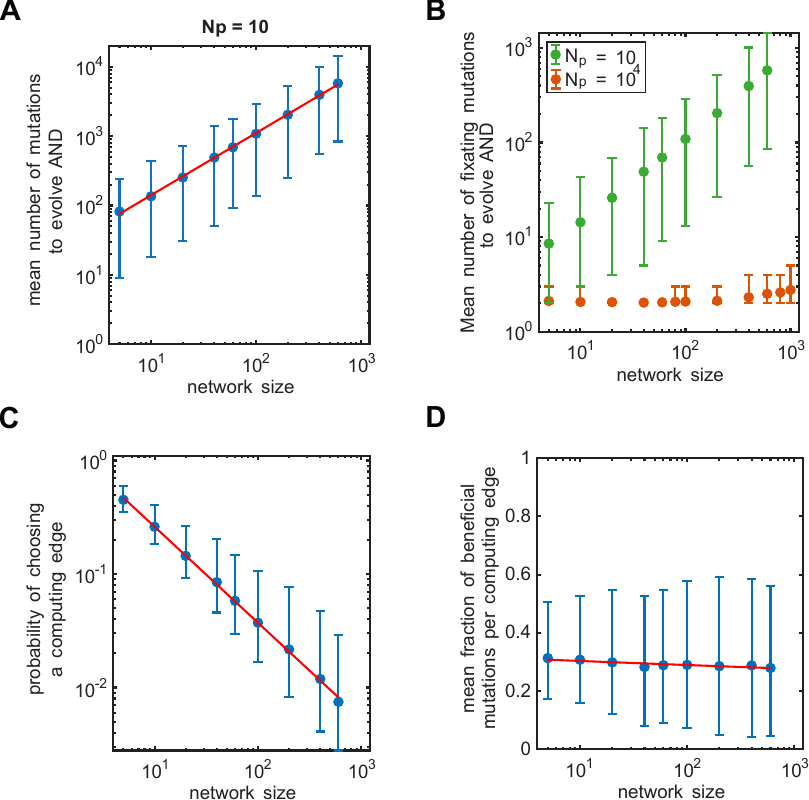}

\caption{Mechanism underlying the scaling of the number of mutations required to evolve AND at $N_P=10$. Simulations are initialized with the network shown in fig. \ref{fig:fig3}A. Dots represent mean statistic across simulation repeats, and error bars indicate the 5th--95th percentile range (see Methods). 
\textbf{(A)} The total number of mutations (counting both fixating and non-fixating) to evolve AND as a function of network size. Red line indicates a power-law fit to all data points with a scaling exponent of $0.90$. Fitting two non-overlapping subsets produced scaling exponents of $0.87$ and $0.92$ for size range 5--40 and 60--600, respectively.
\textbf{(B)} Mean number of fixating mutations during evolution as a function of network size for $N_P=10$, with $N_P=10^4$ shown for comparison. 
\textbf{(C)} Mean probability of choosing a computing edge to mutate, $P(c)$, as a function of network size. Red lines show power-law fit to all data points with a scaling exponent of $-0.85$. Fitting two non-overlapping subsets produced scaling exponents of $-0.81$ and $-0.87$ for size ranges 5--40 and 60--600, respectively.
\textbf{(D)} Mean fraction of beneficial mutations per computing edge, $P(b|c)$, as a function of network size. Red line indicates a power-law fit to all data points with a scaling exponent of $-0.020$. Fitting two non-overlapping subsets produced scaling exponents of $-0.049$ and $-0.012$ for size ranges 5--40 and 60--600, respectively.}
\label{fig:figS4}
\end{figure}
\FloatBarrier

\begin{figure}[htbp]
   \centering
   \includegraphics[width=0.8\textwidth]{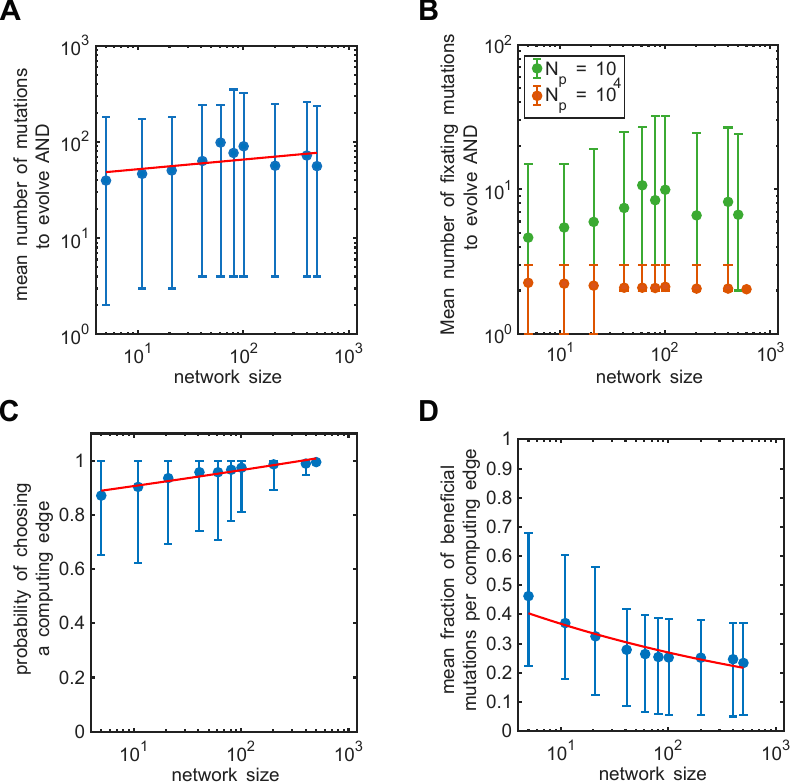}
  
\caption{Effect of network size on the number of mutations needed to evolve target function AND, for simulations initialized by the network shown in fig. \ref{fig:fig5}A with $N_P$ = 10. Dots and error-bars represent mean statistic and 5--95th percentile range from simulation repeats respectively. \textbf{(A)} Mean number of mutations as a function of network size. Red line indicates a power-law fit to all data points, $y=41.3x^{0.10}$. Fits to two non-overlapping subsets yield scaling exponents of $0.21$ and $-0.17$ for size ranges 5--41 and 81--501, respectively. \textbf{(B)} Mean number of fixating mutations during evolution as a function of network size for $N_P=10$, with $N_P=10^4$ shown for comparison. \textbf{(C)} Probability that a mutation targets a computing edge during evolution as a function of network size. Power-law fits across all network sizes, sizes 5--41, and sizes 81--501 yielded scaling exponents of $0.027$, $0.046$, and $0.014$, respectively. \textbf{(D)} Mean fraction of beneficial mutations available per computing edge as a function of network size. Power-law fits across all network sizes, sizes 5--41, and sizes 81--501 yielded scaling exponents of $-0.135$, $-0.238$, and $-0.036$, respectively. The lack of pronounced scaling in panel A is consistent with the lack of scaling in panels C and D at large $N$, as per Equation \ref{eq2}.}
\label{fig:figS5}
\end{figure}
\FloatBarrier

\begin{figure}[htbp]
  \centering
  \includegraphics[scale=0.85]{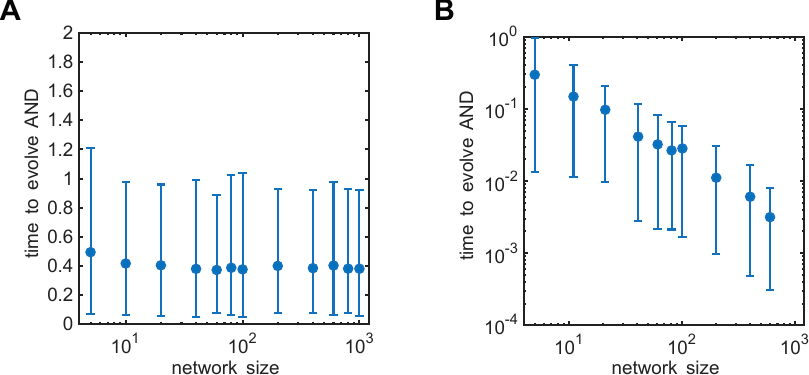}
  \caption{Mean time required to evolve AND as a function of network size for the two initial network configurations. Simulations are performed using $N_P=10^4$. Dots represent the average time across independent simulations at each network size and error bars indicate the 5th–95th percentile range. \textbf{(A)} Simulations initialized with the network shown in fig. \ref{fig:fig3}A. \textbf{(B)} Simulations initialized with the network shown in fig. \ref{fig:fig5}A.}
  \label{fig:figS6}
\end{figure}
\FloatBarrier
\end{document}